\documentclass{kluwer}
\usepackage{graphicx}

\begin{document}
\begin{article}
\begin{opening}
\title{Towards a population of HMXB/NS microquasars as counterparts of low-latitude unidentified EGRET sources}
%\subtitle{Basic Instructions}

\author{M. \surname{Rib\'o}\email{mribo@discovery.saclay.cea.fr}}
\institute{Service d'Astrophysique, CEA Saclay}
\author{J.A. \surname{Combi}}
\institute{Universidad de Ja\'en\\
Instituto Argentino de Radioastronom\'{\i}a (IAR)}
\author{I.F. \surname{Mirabel}}
\institute{Service d'Astrophysique, CEA Saclay\\
Instituto de Astronom\'{\i}a y F\'{\i}sica del Espacio (IAFE)}

%\date: rather not

%\dedication{To Jim}

%\translation{De Kluwer LaTeX stylefile; aanwijzingen voor auteurs}

\runningtitle{HMXB/NS microquasars and unidentified EGRET sources}
\runningauthor{Rib\'o et al.}

\begin{ao}
Marc Rib\'o\\
Service d'Astrophysique\\
CEA Saclay\\
B\^at. 709, L'Orme des Merisiers\\
F-91191 Gif-sur-Yvette, Cedex\\
France
\end{ao} 

%\begin{motto}
%What can't be done with TeX isn't worth doing.
%\end{motto}

\begin{abstract} 
The discovery of the microquasar LS~5039 well within the 95\% conficence
contour of the Unidentified EGRET Source (UES) 3EG~J1824$-$1514 was a major
step towards the possible association between microquasars (MQs) and UESs. The
recent discovery of precessing relativistic radio jets in LS~I~+61~303, a
source associated for long time with 2CG~135+01 and with the UES
3EG~J0241+6103, has given further support to this idea. Finally, the very
recently proposed association between the microquasar candidate
AX~J1639.0$-$4642 and the UES 3EG~J1639$-$4702 points towards a population of
High Mass X-ray Binary (HMXB)/Neutron Star (NS) microquasars as counterparts
of low-latitude unidentified EGRET sources. 
\end{abstract}


%\abbreviations{\abbrev{KAP}{Kluwer Academic Publishers};
%\abbrev{compuscript}{Electronically submitted article}}

%\nomenclature{\nomen{KAP}{Kluwer Academic Publishers};
%\nomen{compuscript}{Electronically submitted article}}

%\classification{JEL codes}{D24, L60, 047}
\end{opening}

\section{Introduction}

The third EGRET catalog (Hartman et~al. \citeyear{hartman99}) contains 271
point sources detected at energies above 100~MeV. The majority of these
sources, $\sim$168 or $\sim$62\%, still remain unidentified. Among them, there
are 72 sources located at low galactic latitudes, having $|b|$$<$10$^{\circ}$,
which represents around 45\% of the UES population. Therefore, several of
these objects are presumably of galactic nature. Similar properties between
some of these UESs, indicate that there are at least three different groups of
galactic populations (Romero et~al. \citeyear{romero04}, Grenier
\citeyear{grenier04}). The group of young stellar objects and star-forming
regions (Romero \citeyear{romero01}), those sources forming a halo around the
galactic center and a group of sources correlated with the Gould Belt (Grenier
\citeyear{grenier00}). 

Based both on multiwavelength observations and theory, microquasars (see
Mirabel \& Rodr\'{\i}guez \citeyear{mirabel99} for a review) with massive
companions have been proposed as possible counterparts of the first group of
galactic UESs by several authors (Paredes et~al. \citeyear{paredes00}, Kaufman
Bernad\'o et~al. \citeyear{kaufman02}, Romero et~al. \citeyear{romero04},
Bosch-Ramon et~al. \citeyear{bosch04}). In Sects.~2 and 3 of this paper we
will briefly review the properties of the two well-known microquasars LS~5039
and LS~I~+61~303, typically associated with the first group of UESs, while in
Sect.~4 we will present the possible association between the microquasar
candidate AX~J1639.0$-$4642 and the UES 3EG~J1639$-$4702. Finally, in Sect.~5
we will compare the available data of these 3 sources from radio to
gamma-rays, and we will discuss on similarities pointing towards a population
of HMXB with NS microquasars as counterparts of low-latitude UESs.

\section{LS~5039}

The high mass X-ray binary system LS~5039 (Paredes et~al.
\citeyear{paredes00}) is one of the $\sim$15 confirmed galactic microquasars
(Rib\'o \citeyear{ribo04}). LS~5039 is a bright $V$$\sim$11.2 star with an
ON6.5V((f)) spectral type (McSwain et~al. \citeyear{mcswain04}). The binary
system has a short orbital period of $P=4.4267 \pm0.0005$~d, a high
eccentricity of $e=0.48\pm0.06$, and a low mass function
$f(m)=0.0017\pm0.0005$~$M_{\odot}$, suggesting the presence of a NS as the
compact object in the system (McSwain et~al. \citeyear{mcswain04}).

%------------------------------------------------------------------------------
\begin{figure}[t!] %figure 1
\center
\resizebox{1.0\hsize}{!}{\includegraphics[angle=0]{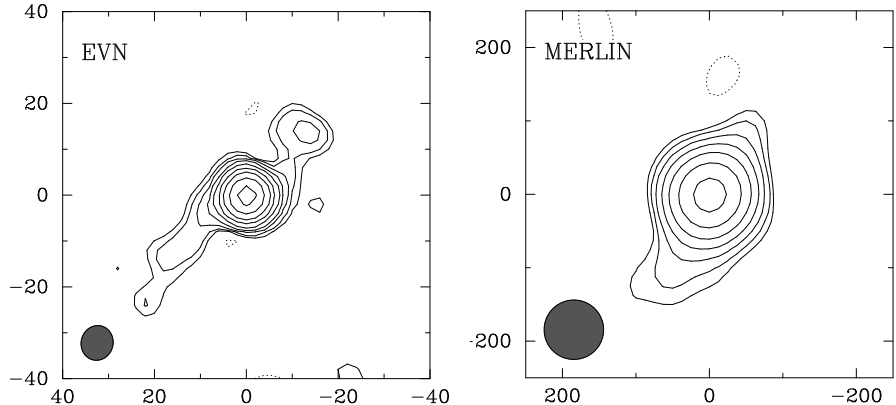}}
\caption{Left: EVN self-calibrated image of LS~5039 at 5~GHz obtained on 2000
March~1. Axes units are in mas. The synthesized beam is plotted in the lower
left corner.
%, has a size of 7.60$\times$6.96~mas in P.A.\ of $-14^{\circ}$. The first
%contour corresponds to 0.3~mJy~beam$^{-1}$, while consecutive ones scale with
%$3^{1/2}$. 
Right: Same as left but obtained with MERLIN.
%The convolving circular beam has a diameter of 81~mas. The first contour
%corresponds to a flux density of 1~mJy~beam$^{-1}$, while consecutive ones
%scale as well with $3^{1/2}$. 
From Paredes et~al. (2002).}
\label{ls5039_evn_merlin}
\end{figure}
%------------------------------------------------------------------------------

Observations conducted with the EVN and MERLIN (see
Fig.~\ref{ls5039_evn_merlin}) confirmed the persistent nature of this MQ, and
revealed the presence of an asymmetric two-sided jet reaching up to 1000~AU on
the longest jet arm (Paredes et~al. \citeyear{paredes02}). These observations
also suggest a bending of the jets with increasing distance from the core
and/or precession.

\begin{figure}[t!] %figure 2
%\vspace{10cm}
%\special{psfile=3EG_J1824-1514.eps hscale=20 vscale=20 hoffset=-15 voffset=-100 angle=0}
\center
\resizebox{0.65\hsize}{!}{\includegraphics[angle=0]{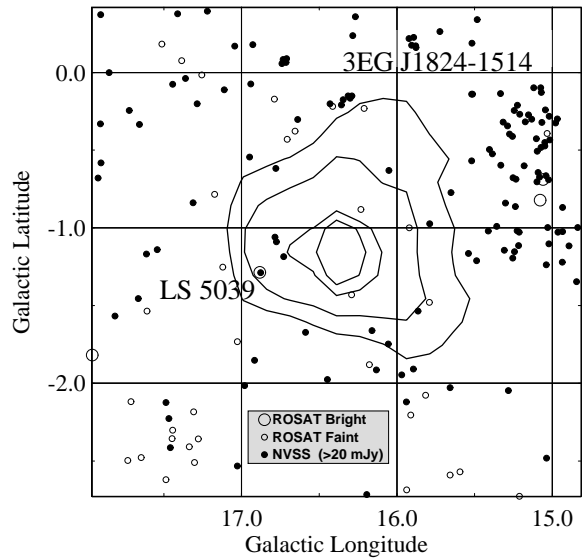}}
\caption[]{Location map of 3EG~J1824$-$1514. The contours represent, from inside to outside, the 50\%, 68\%, 95\%, and 99\% statistical probability that a $\gamma$-ray source lies within the given contour.
%The open big circles are ROSAT bright sources, the small open circles are the
%ROSAT faint sources and the filled circles are radio sources from the NVSS
%survey. 
The only source with X-ray and radio emission (filled circle inside an open big circle, $l=16.88^{\circ}$ and $b=-1.29^{\circ}$), and well inside the 95\% contour, is LS~5039 (Rib\'o \citeyear{ribo02}).}
\label{3egj1824}
\end{figure}

The possibility that LS~5039 is a $\gamma$-ray emitter was suggested by
Paredes et~al. (\citeyear{paredes00}), who proposed the association of the
system with the UES 3EG~J1824$-$1514 (Hartman et~al. \citeyear{hartman99}). We
show in Fig.~\ref{3egj1824} the location map of the $\gamma$-ray source
together with the NVSS and bright/faint ROSAT sources. The only simultaneous
X-ray/radio source within the statistical contours of 3EG~J1824$-$1514 is the
microquasar LS~5039. We note that this binary system is present in the BATSE
Earth occultation catalog of low-energy gamma-ray sources (Harmon et~al.
\citeyear{harmon04}), with a positive detection of a few mCrab up to
$\sim$100~keV. The source is not present in cumulative observations conducted
with the INTEGRAL satellite (Bird et~al. \citeyear{bird04}), although it is
expected to be detected when adding a few more months of data. We also point
out that there is an unidentified COMPTEL source with a position compatible
with LS~5039 (Collmar \citeyear{collmar04}).

%------------------------------------------------------------------------------
%------------------------------------------------------------------------------

Astrometric studies carried out by Rib\'o et~al. (\citeyear{ribo02}), show
that it is a runaway system with a systemic velocity of $\sim$150 km~s$^{-1}$
that moves away from the Galactic plane with a velocity of $\sim$100
km~s$^{-1}$. This result, combined with the possible lifetime of the donor
star, indicates that it could reach a not-so-low galactic latitude of
$b=-12^{\circ}$ still behaving as a microquasar.

Bosch-Ramon \& Paredes (\citeyear{bosch04a}) have recently developed a
detailed numerical model to test whether this system can actually produce the
emission detected by EGRET through inverse Compton (IC) scattering. Their
numerical approach considers a population of relativistic electrons entrained
in a cylindrical inhomogeneous jet, which interact with both the radiation and
the magnetic fields. The computed spectrum is able to reproduce the observed
spectral characteristics at very high (GeV) energies.

\section{LS~I~+61~303}

The Be/X-ray binary system LS~I~+61~303 is a well-studied object since it
presents radio and X-ray variability linked to its $\sim$26.5~d orbital period
(Gregory \citeyear{gregory02}; Paredes et~al. \citeyear{paredes97}). The donor
star in this system is a rapidly rotating B0V star with variable mass loss
(Hutchings \& Crampton \citeyear{hutchings81}). Some properties of this system
can be explained assuming that the unseen companion is a non-accreting young
pulsar with a relativistic wind strongly interacting with the wind of the Be
star (Maraschi \& Treves \citeyear{maraschi81}). On the contrary, other
properties of LS~I~+61~303 fit better a model where the companion is accreting
even with two episodes of super-critical accretion along the orbit (Mart\'{\i}
\& Paredes \citeyear{marti95}).

%------------------------------------------------------------------------------
\begin{figure}[t!] %figura 3
%\vspace{9.3cm}
%\special{psfile=3EG_J0241+6103.eps hscale=50 vscale=50 hoffset=20 voffset=-80 angle=0}
\center
\resizebox{0.55\hsize}{!}{\includegraphics[angle=0]{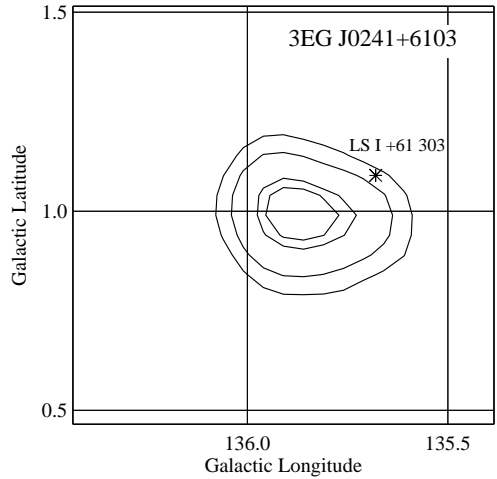}}
\caption[]{Location map of 3EG~J0241+6103. The position of LS~I~+61~303 is marked with an asterisk. Adapted from Hartman et~al. (\citeyear{hartman99}).}
\label{3egj0241}
\end{figure}
%------------------------------------------------------------------------------

\begin{figure}[t!] %figura 4
\center
\resizebox{1.0\hsize}{!}{\includegraphics[angle=0]{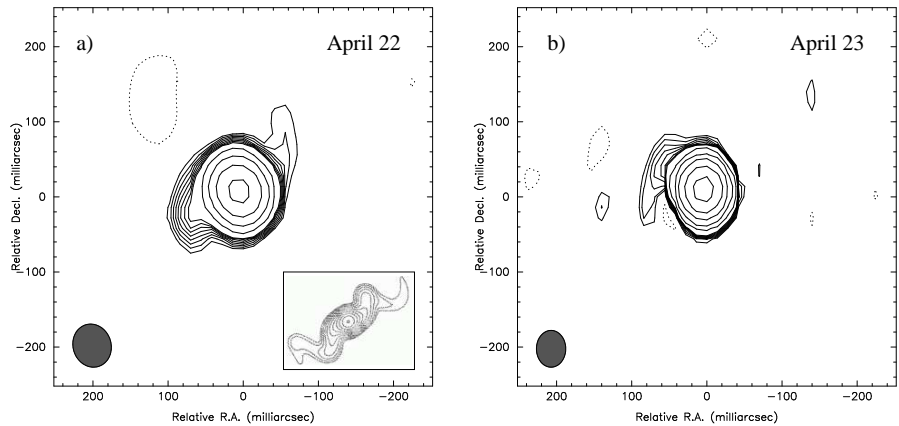}}
\caption[]{
a) MERLIN self-calibrated image of LS~I~+61~303 at 5~GHz and using
natural weights, obtained on 2001 April 22.
%North is up and East is to the left.
The synthesized beam is plotted in the lower left corner.
%has a size of $51\times58$~mas, with a PA of 17$^{\circ}$. The contour levels
%are at $-$3, 3, 4, 5, 6, 7, 8, 9, 10, 20, 40, 80, and 160$\sigma$, being
%$\sigma$=0.14~mJy~beam$^{-1}$. 
The S-shaped morphology strongly recalls the precessing jet of SS~433, whose
simulated radio emission (Fig.~6b in Hjellming \& Johnston
\citeyear{hjellming88}, rotated here for comparison purposes) is given in the
small box. 
b) Same as before but for the April 23 run and using uniform weights for better displaying.
%The synthesized beam has a size of $39\times49$~mas, with a PA of
%$-$10$^{\circ}$. The contour levels are the same as those used in the April 22
%image but up to 320$\sigma$, with $\sigma$=0.12~mJy~beam$^{-1}$. 
From Massi et~al. (\citeyear{massietal04}).}
\label{lsi_merlin}
\end{figure}

This X-ray binary system has been associated for long time with the gamma-ray
source 2CG~135+01/3EG~J0241$+$6103 (see Fig.~\ref{3egj0241}), which displays
variability on timescales of days (Tavani et~al. \citeyear{tavani96},
\citeyear{tavani98}; Wallace et~al. \citeyear{wallace00}). During the last
years, Massi et~al. (\citeyear{massi01}, \citeyear{massietal04}) have revealed
its MQ nature through the discovery of a radio jet (see Fig.~\ref{lsi_merlin})
extending 200~AU at both sides of a central core, that appears to experience a
fast precession, which could explain the short-term gamma-ray variability of
3EG~J0241$+$6103 (as proposed by Kaufman Bernad\'o et~al.
\citeyear{kaufman02}) and the puzzling VLBI structures found in previous
observations. This result points to the occurrence of accretion/ejection
processes in this system, ruling out, in principle, the non-accreting young
pulsar scenario.

%------------------------------------------------------------------------------
%------------------------------------------------------------------------------

Massi (\citeyear{massi04}) has recently studied the data acquired within the
pointed EGRET observations of 3EG~J0241$+$6103 and claimed the detection of a
periodicity of $P=27.4\pm7.2$~d, consistent with the orbital period of the
binary system. If this is confirmed, the identification of LS~I~+61~303 as the
counterpart of 3EG~J0241$+$6103 would be unambiguous. In any case, an
important point is that all the available data are compatible with an increase
of $\gamma$-ray emission around periastron, that can be tested with Cherenkov
telescopes and future satellites.

This binary system is also present in the BATSE Earth occultation catalog of
low-energy gamma-ray sources (Harmon et~al. \citeyear{harmon04}), with a
positive detection of a few mCrab up to $\sim$100~keV, although the detection
is not as significant as in the case of LS~5039. LS~I~+61~303 is not present
in cumulative observations conducted with the INTEGRAL satellite (Bird et~al.
\citeyear{bird04}), although it is expected to be detected when adding a few
more months of data, like in the case of LS~5039. We note that there is a
COMPTEL source containing LS~I~+61~303 and the quasar QSO~0241+622 (van Dijk
et~al. \citeyear{vandijk96}).

A numerical model to explain the EGRET emission of LS~I~+61~303 has also been
developed by Bosch-Ramon \& Paredes (\citeyear{bosch04b}).

\section{AX~J1639.0$-$4642}

Aimed at discovering new MQs, Combi et~al. (\citeyear{combi04}) have recently
carried out a multiwavelength study of the unidentified X-ray source
AX~J1639.0$-$4642. This object was discovered by the Advanced Satellite for
Cosmology and Astrophysics (ASCA) observatory at the 0.7--10~keV energy range,
and presented as a possible HMXB (Sugizaki et~al. \citeyear{sugizaki01}). Its
measured flux was $F_{\rm X (0.7-10~keV)}$= $19.2 \times 10^{-12}$
erg~cm$^{-2}$~s$^{-1}$, and it showed variable X-ray emission, with a
confidence $\geq$ 99 \%. Its spectrum was fitted with a power law with a very
hard photon index $\Gamma=-0.01^{+0.66}_{-0.60}$ and a poorly constrained
hydrogen column density of $N_{\rm
H}=12.82^{+8.58}_{-6.88}\times10^{22}~$cm$^{-2}$. Combi et~al.
(\citeyear{combi04}) re-analyzed these data and found evidences for
variability on timescales of hours.

%------------------------------------------------------------------------------
\begin{figure}[t!] %figura 5
\center
\resizebox{0.65\hsize}{!}{\includegraphics[angle=0]{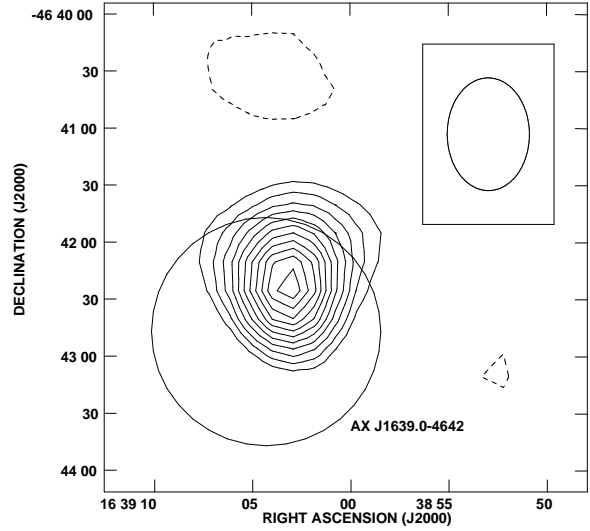}}
\caption[]{Contour image of the MGPS data obtained with MOST at 843~MHz on 1992 April 16.
%The image size is 4$^{\prime}$$\times$4$^{\prime}$.
The radio source MOST~J1639.0$-$4642 is well within the 90\% uncertainty error circle of the X-ray source AX~J1639.0$-$4642.
%Contours are $-$2, 2, 3, 4, 5, 6, 7, 8, 9, 10, 11 and 12 times the rms noise
%level of 10~mJy. 
The ellipse in the top right corner is the convolving beam.
%of 59.2$\times$43.0 arcsec in PA=0$^{\circ}$. 
From Combi et~al. (\citeyear{combi04}).}
\label{most}
\end{figure}
%------------------------------------------------------------------------------

%------------------------------------------------------------------------------
\begin{figure}[t!] %figura 6
\center
\resizebox{0.65\hsize}{!}{\includegraphics[angle=0]{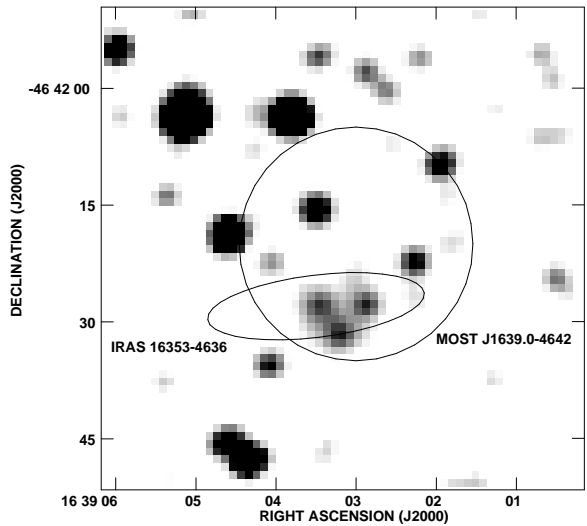}}
\caption[]{2MASS $K_{\rm s}$-band %60$^{\prime\prime}$$\times$60$^{\prime\prime}$ 
image of the environment of the radio source MOST~J1639.0$-$4642, whose
3$\sigma$ position error circle is shown, together with the 2$\sigma$ position
error ellipse of the far infrared source IRAS~16353$-$4636. Several NIR
sources from the 2MASS catalog are possible counterparts of both sources. From
Combi et~al. (\citeyear{combi04}).}
\label{2mass}
\end{figure}
%------------------------------------------------------------------------------

In searching for radio sources in the field of AX~J1639.0$-$4642, Combi et~al.
(\citeyear{combi04}) found that the Molonglo Galactic Plane Survey (MGPS) at
843~MHz (Green et~al. \citeyear{green99}) revealed a point-like radio source
(see Fig.~\ref{most}), dubbed MOST~J1639.0$-$4642, well within the error box
of the X-ray source, with a flux density of $136\pm18$~mJy. At near infrared
(NIR) wavelengths Combi et~al. (\citeyear{combi04}) inspected the 2~Micron All
Sky Survey (2MASS, Cutri et~al. \citeyear{cutri03}), and found 10 sources in
the 3$\sigma$ error circle in position of MOST~J1639.0$-$4642, some of them
visible in the $K_s$-band image shown in Fig.~\ref{2mass}.
%At optical wavelengths they searched in the USNO-B1.0 catalog (Monet et~al.
%\citeyear{monet03}), and found 6 sources in the same error circle, of which 3
%were located within 1 arcsecond of 2MASS sources.
At the far infrared part of the spectrum, from 12 to 100 microns, they found
that the source IRAS~16353$-$4636 lies inside the error box of the X-ray
source.
%This infrared source is located at $(l, b) = (337\fdg995, +0\fdg077)$,
%$(\alpha, \delta)_{\rm J2000.0} = (16^{\rm h} 39^{\rm m} 03\fs5, -46\degr
%42\arcmin 28\arcsec)$ (95\% or 2$\sigma$ uncertainty ellipse of
%14\arcsec$\times$4\arcsec\ in PA=97\degr). The infrared fluxes at 12, 25, 60
%and 100 microns are 12.5$\pm$0.6, 73.3$\pm$3.7, $<$806 (3$\sigma$ upper limit)
%and 2230$\pm$330~Jy, respectively. After correction for the slope of the
%spectrum, these fluxes become 13.5$\pm$0.7, 80.5$\pm$4.0, $<$806 (3$\sigma$
%upper limit) and 2210$\pm$330~Jy, respectively. 
This source overlaps the southern part of the 3$\sigma$ position error circle
of MOST~J1639.0$-$4642, and its location uncertainty ellipse contains
several 2MASS sources, as can be seen in Fig.~\ref{2mass}. The X-ray source
AX~J1639.0$-$4642 has been recently re-discovered at higher energies with the
IBIS telescope onboard the INTEGRAL satellite, dubbed IGR~J16393$-$4643
(Malizia et~al. \citeyear{malizia04}; Bird et~al. \citeyear{bird04}).
%It is located at $(l, b) = (338\fdg02, +0\fdg04)$, $(\alpha, \delta)_{\rm
%J2000.0} = (16^{\rm h} 39^{\rm m} 18^{\rm s}, -46\degr 43\arcmin 00\arcsec)$
%(90\% or 1.6$\sigma$ uncertainty of 2\arcmin). We note that although this
%error circle does not include either the center of \object{AX~J1639.0$-$4642}
%or the source MOST~J1639.0$-$4642, both sources fall within the 2$\sigma$
%error circle of \object{IGR~J16393$-$4643}. The new position by Bird et~al.
%\citeyear{bird04} does include the radio source.
This source shows an average flux of $F_{\rm X (20-100~keV)}\simeq5\times
10^{-11}$ erg~cm$^{-2}$~s$^{-1}$, and presents a factor of 2--3 flux
variability on timescales of months.

Although there is no spectroscopic/photometric information of a NIR/optical
counterpart to derive a distance to the source, assuming that it is located in
the Scutum-Crux or in the Norma spiral arms, a range of distances between 3
and 13~kpc is obtained.

%------------------------------------------------------------------------------
\begin{figure}[t!] %figure 7
%\vspace{7.6cm}
%\special{psfile=3EG_J1639-4702.eps hscale=55 vscale=55 hoffset=0 voffset=-110 angle=0}
\center
\resizebox{0.65\hsize}{!}{\includegraphics[angle=0]{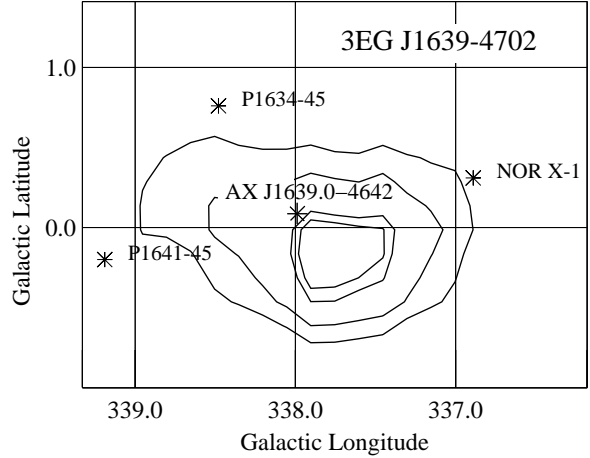}}
\caption[]{Location map of 3EG~J1639$-$4702. The position of the X-ray source AX~J1639.0$-$4642, well within the 95\% probability contour, is marked with an asterisk. Other sources appearing outside the 99\% probability contour are also indicated. From Combi et~al. (\citeyear{combi04}).}
\label{3egj1639}
\end{figure}
%------------------------------------------------------------------------------

Combi et~al. (\citeyear{combi04}) pointed out that AX~J1639.0$-$4642 lies
inside the 95\% location contour of the UES 3EG~J1639$-$4702 (Hartman et~al.
\citeyear{hartman99}) as can be seen in Fig.~\ref{3egj1639}. Its $\gamma$-ray
flux is $(53.2\pm8.7)\times10^{-8}$~photon~cm$^{-2}$~s$^{-1}$, presents a
steep $\gamma$-ray spectral index of $\Gamma=2.5\pm0.18$ and has a variability
index of $I = 1.95$. Although Torres et~al. (\citeyear{torres01a}) found three
radio pulsars inside the 95\% confidence contour of the $\gamma$-ray source,
its possible variability and steep photon index do not seem to agree, in
principle, with a pulsar origin. Similarly, these properties would rule out an
association with the three SNRs found within the 95\% confidence contour
(Torres et~al. \citeyear{torres03}). Moreover, no identified blazar has been
found within the $\gamma$-ray contours. Therefore, Combi et~al.
(\citeyear{combi04}) suggested that the microquasar candidate
AX~J1639.0$-$4642/MOST~J1639.0$-$4642 (=IGR~J16393$-$4643) is the counterpart
of 3EG~J1639$-$4702. Observations with ATCA are in progress to unveil the
nature of this source.

\section{HMXB/NS microquasars as counterparts of low-latitude unidentified EGRET sources}

As discussed above, the possibility of MQs being $\gamma$-ray emitters was
suggested by Paredes et~al. (\citeyear{paredes00}), who proposed the
association between the HMXB LS~5039 and the UES 3EG~J1824$-$1514. In their
scenario (Paredes et~al. \citeyear{paredes00}, \citeyear{paredes02}) the
$\gamma$-rays are produced by IC upscattering of stellar ultraviolet (UV)
photons by the non-thermal relativistic electron population that later on will
produce the detected radio emission. Recently, more detailed models
considering precession (Kaufman Bernad\'o et~al. \citeyear{kaufman02}),
hadronic jets in windy microquasars (Romero et~al. \citeyear{romero03}) and
all possible photon fields (Bosch-Ramon et~al. \citeyear{bosch04}) have been
proposed to explain the high-energy gamma-ray emission from HMXB microquasars.

On the other hand, as already stated, the X-ray binary system LS~I~+61~303 has
been associated with the UES 3EG~J0241$+$6103, and Massi et~al.
(\citeyear{massi01}, \citeyear{massietal04}) have revealed its MQ nature. If
the MQ nature of AX~J1639.0$-$4642 is confirmed, it could be the third MQ
source related to a UES.

%------------------------------------------------------------------------------
\begin{table} %Table 1
\tiny
\begin{tabular}{ccccc}
\hline
$\gamma$-ray source& $\Gamma_{\gamma}$$^{a}$ & $I$\,$^{b}$ & $L_{\gamma (>100~{\rm MeV})}$$^{c}$ & $d$ \\
                   &                         &             & (erg~s$^{-1}$)                      & (kpc)\\
\hline
3EG~J1824$-$1514 & 2.19$\pm$0.18 & 3.00 & ~~~~\,3.6$\times$10$^{35}$     & \,2.9\,$^{d}$\\
3EG~J0241$+$6103 & 2.21$\pm$0.07 & 1.31 & ~~~~\,3.1$\times$10$^{35}$     & 2.0\,$^{e}$\\
3EG~J1639$-$4702 & 2.50$\pm$0.18 & 1.95 & (3--50)$\times$10$^{35}$ & 3--13 ?\\
\hline
\end{tabular}
\caption[]{Properties of the three $\gamma$-ray sources discussed in the text. The luminosity interval for AX~J1639.0$-$4642 reflects the range of assumed possible distances.}
\label{table:egret}
\tiny{$^{a}$ Hartman et~al. (\citeyear{hartman99}); $^{b}$ Torres et~al. (\citeyear{torres01b}); $^{c}$ computed using the\\
photon fluxes and indices from Hartman et~al. (\citeyear{hartman99}); $^d$  Rib\'o et~al. (\citeyear{ribo02});\\ 
$^{e}$ Frail \& Hjellming (\citeyear{frail91}).} 
\end{table}
%------------------------------------------------------------------------------

%------------------------------------------------------------------------------
\begin{table} %Tabel 2
\tiny
\begin{tabular}{cccccc}
\hline
X-ray source  & $L_{\rm X (0.7-10~keV)}$ & $L_{\rm radio (0.1-100~GHz)}$  &  Spectral   & $P_{\rm orb}$  & $d$ \\
              &  (erg~s$^{-1}$)          & (erg~s$^{-1}$)                 & type        & (days)         & (kpc)\\
\hline
LS~5039 & (0.5--5)$\times$10$^{34}$\,$^{a}$      & ~~~~$\sim$1.0$\times$10$^{31}$\,$^{b}$  & ON6.5V((f))\,$^{c}$ & ~4.4\,$^{c}$  & \,2.9\,$^{d}$\\
LS~I~+61~303 & ~~(1--6)$\times$10$^{34}$\,$^{e}$ & ~~\,(1--17)$\times$10$^{31}$\,$^{f}$ & B0Ve\,$^{g}$        &  26.5\,$^{h}$  & 2.0\,$^{i}$ \\
AX~J1639.0$-$4642 & (2--40)$\times$10$^{34}$~\,  & (0.8--16)$\times$10$^{31}$\,$^{j}$  & ?                   & ?              & 3--13 ?\\
\hline
\end{tabular}
\caption[]{Proposed X-ray/optical/radio counterparts of the three $\gamma$-ray sources of Table~\ref{table:egret}. In the cases of the microquasars LS~5039 and LS~I~+61~303 the luminosity intervals correspond to intrinsic variability of the sources at the corresponding distances, while in the case of AX~J1639.0$-$4642 they reflect the range of assumed possible distances.}
\label{table:xray}
\tiny{$^a$ Reig et~al. (\citeyear{reig03}); $^b$ Mart\'{\i} et~al. (\citeyear{marti98}), Rib\'o et~al. (\citeyear{ribo99}); $^c$ McSwain et~al. (\citeyear{mcswain04}); $^d$ Rib\'o et~al. (\citeyear{ribo02}); $^e$ Paredes et~al. (\citeyear{paredes97}); $^f$ computed from the values quoted in Strickman et~al. (\citeyear{strickman98}); $^g$ Hutchings \& Crampton (\citeyear{hutchings81}); $^h$ Gregory (\citeyear{gregory02}); $^{i}$ Frail \& Hjellming (\citeyear{frail91}); $^j$ computed assuming $\alpha=-1$.}
\end{table}
%-----------------------------------------------------------------------------

We quote the basic properties of these three $\gamma$-ray sources in
Table~\ref{table:egret}, and the properties of the proposed X-ray counterparts
in Table~\ref{table:xray}. There are 3 observational facts that should be
noted. The first one is that LS~5039 and LS~I~+61~303, and probably
AX~J1639.0$-$4642, have massive optical companions, which provide an intense
stellar UV photon field. On the other hand, the compact object appears to be
compatible with a neutron star in the cases of LS~5039 (McSwain et~al.
\citeyear{mcswain04}) and LS~I~+61~303 (Hutchings \& Crampton
\citeyear{hutchings81}; Casares et~al, \citeyear{casares04}; but see Massi
\citeyear{massi04}). Finally, it is interesting to point out that the
luminosities obtained in each spectral domain are very similar in all three
sources, specially for the shorter distances to AX~J1639.0$-$4642, giving
support to the idea that all of them have similar emission processes.

Moreover, it should be noted that LS~5039 and LS~I~+61~303 are the only
microquasars having both a high-mass donor and (possibly) a NS as the compact
object, and that there are no other microquasars (containing low-mass donors
and/or black holes) located within the probability contours of unidentified
EGRET sources. Therefore, a strong statement can be made: HMXB/NS microquasars
appear as good counterparts of low-latitude unidentified EGRET sources.

This statement is followed by some natural questions. 1) May LS~5039 and
LS~I~+61~303 still contain black holes (BHs)? Although formally possible, this
is unlikely, because regarding the properties of their X-ray emission they
should be in the so-called low/hard state (Fender \& Maccarone
\citeyear{fender04}), but they do not follow the empirical correlation between
X-ray and radio flux found by Gallo et~al. (\citeyear{gallo03}), in the sense
that they are clearly too radio loud. 2) Why no HMXB microquasars with BHs
(e.g., Cygnus~X-1) are present in the third EGRET catalog? A possibility is
that black hole state changes may play a role, preventing the detection of
gamma-ray emission when the jet is not present during the high/soft state (but
this state is very rare in the case of Cygnus~X-1 and in fact, as
discussed by Romero et~al. (\citeyear{romero02}), a low high-energy cutoff of
a few hundreds in the Lorentz factor of the electrons is compatible with the
data). 3) Why no Low Mass X-ray Binary (LMXB) microquasars are present in the
third EGRET catalog? One possibility is that the optical companions do not
provide the necessary intense UV radiation fields needed for an effective IC
process to produce high-energy gamma-rays. This possibility is strongly model
dependent, and not valid if Self Synchrotron Compton losses are dominant. The
other possibility is that LMXBs are in general terms transient objects, that
maybe were not active during the EGRET viewing periods, while HMXBs tend to be
persistent systems. 4) Of course, BH and LMXB microquasars could still emit
high-energy gamma-rays and not be present in the third EGRET catalog because
of the relatively poor sensitivity threshold of the instrument.

In conclusion, persistent HMXBs containing NSs not experiencing state changes
are good candidates for the counterparts of the still unidentified high-energy
gamma-ray sources at low-galactic latitudes (approximately up to
$|b|<10^{\circ}$, as discussed for LS~5039), and we consider that these
objects may define a population among UESs. Observations with the future
missions AGILE and GLAST will confirm or reject both the proposed associations
between these microquasars and the corresponding high-energy gamma-ray
sources, and the hypothesis discussed above.

\acknowledgements
We thank Sylvain Chaty, Paula Benaglia, Gustavo E. Romero, Josep M. Paredes,
Josep Mart\'{\i}, Valent\'{\i} Bosch-Ramon and Rob Fender for useful
discussions, and an anonymous referee for useful comments that helped to improve the paper.
M.R. acknowledges support by a Marie Curie Fellowship of the European Community programme Improving Human Potential under contract number HPMF-CT-2002-02053. M.R. also acknowledges partial support by DGI of the Ministerio de Ciencia y Tecnolog\'{\i}a (Spain) under grant AYA2001-3092, as well as partial support by the European Regional Development Fund (ERDF/FEDER).
J.A.C. is a researcher of the programme {\em Ram\'on y Cajal} funded by the Spanish Ministery of Science and Technology and the University of Ja\'en. J.A.C. was supported by CONICET (under grant PEI 6384/03). 

%This research has made use of the NASA's Astrophysics Data System Abstract
%Service, of the SIMBAD database, operated at CDS, Strasbourg, France, and of
%the NASA/IPAC Extragalactic Database (NED) which is operated by the Jet
%Propulsion Laboratory, California Institute of Technology, under contract with
%the National Aeronautics and Space Administration. The Digitized Sky Survey
%was produced at the Space Telescope Science Institute under U.S. Government
%grant NAG~W-2166.
%This publication makes use of data products from the Two Micron All Sky
%Survey, which is a joint project of the University of Massachusetts and the
%Infrared Processing and Analysis Center/California Institute of Technology,
%funded by the National Aeronautics and Space Administration and the National
%Science Foundation.

\end{article}
\end{document}